\documentclass[aps,prb,twocolumn,superscriptaddress,showpacs]{revtex4}
\usepackage{amsmath,amssymb}
\usepackage{graphics}
\usepackage{graphicx}
\usepackage{epsfig}
\usepackage{float}
\usepackage[usenames]{color}
\newcommand{\boldnabla}{\mbox{\boldmath$\nabla$}}

\def\mr{\mathrm}
\def\mc{\mathcal}

\newcommand {\apgt} {\ {\raise-.5ex\hbox{$\buildrel>\over\sim$}}\ }
\newcommand {\aplt} {\ {\raise-.5ex\hbox{$\buildrel<\over\sim$}}\ }


\begin{document}

\title{Doping dependence of fluctuation diamagnetism  in  High $T_\mr{c}$ superconductors}

\author{Kingshuk Sarkar }
\affiliation{Department of Physics, Indian Institute of Science, Bangalore 560012, India}
\author{Sumilan Banerjee}
\affiliation{Department of Condensed Matter Physics, Weizmann Institute of Science, Rehovot 76100, Israel}
\author{Subroto Mukerjee }
\affiliation{Department of Physics, Indian Institute of Science, Bangalore 560012, India}
\author{T.~V.~Ramakrishnan}
\affiliation{Department of Physics, Indian Institute of Science, Bangalore 560012, India}
\affiliation{Department of Physics, Banaras Hindu University, Varanasi 221005, India}

\begin{abstract}
{Using a recently proposed Ginzburg-Landau-like lattice free energy
functional due to Banerjee et~al.~Phys.~Rev.~B 83, 024510 (2011) we
calculate the fluctuation diamagnetism of high-$T_\mr{c}$ superconductors as a function of doping, magnetic field and temperature. 
We analyse the pairing fluctuations above the superconducting transition temperature in the cuprates, ranging from the strong phase fluctuation dominated underdoped limit to the more conventional amplitude fluctuation dominated overdoped regime. We show that a model where the pairing scale increases and the superfluid density decreases with underdoping produces features of the observed magnetization in the pseudogap region, in good qualitative and reasonable quantitative agreement with the experimental data. In particular, we explicitly show that even when the pseudogap has a pairing origin the magnetization actually tracks the superconducting dome instead of the pseudogap temperature, as seen in experiment. We discuss the doping dependence of the `onset' temperature for fluctuation diamagnetism and comment on the role  of vortex core-energy in our model.}
\end{abstract}

\maketitle

\section{Introduction} \label{sec.Introduction}
In recent years  fluctuation diamagnetism and the Nernst effect in the pseudogap regime of cuprate superconductors have generated great interest both theoretically~
\cite{Podolsky,Oganesyan,Mukerjee,Ussishkin,Serbyn,Bolech,Wachtel2014,Jiang2014} and experimentally~\cite{Ong_2003,Romano_2002,Romano_2003,Wang_2005,Wang_2006,Li_2010,Li_2013,Chang_2012,Taillefer_2009}. Experiments have found a very large diamagnetic and Nernst response in the enigmatic pseudogap phase of the cuprates. A large diamagnetic signal naturally points towards fluctuating superconductivity as one of the possible origins. Also, the fact that the Nernst response is usually very small in typical nonmagnetic metals and a much stronger response is observed in the vortex-liquid regime as is expected in a fluctuating superconductor supports this point of view. Compared to conventional superconductors, the Nernst and diamagnetic response have been found to exist over an anomalously large region \cite{Wang_2006,Li_2010} in the pseudogap phase, extending to temperatures far above the superconducting transition temperature $T_\mr{c}$ (see however ref.~\onlinecite{Yu2012,Kokanovic2014}). This, and other mysterious features of the pseudogap phase\cite{Timusk1999}, have lead to an intense debate over whether these responses originate purely from superconducting (SC) fluctuations~\cite{Wang_2005,Wang_2006,Li_2010,Li_2013} or have a significant contribution from quasiparticles and other possible competing orders~\cite{Chang_2012,Taillefer_2009,Daou2010,Hackl2010}. 

Another important piece of the debate is related to the fact that the putative boundary \cite{Wang_2006,Li_2010} of the large Nernst and diamagnetic response regime, the so-called `onset' temperature $T_\mr{onset}$, tracks $T_\mr{c}(x)$ and follows a dome-shaped curve as a function of doping $x$ instead of tracking the pseudogap temperature scale $T^*(x)$, which monotonically decreases with $x$. This has been argued as evidence against a pairing origin of the pseudogap, mainly due to the expectation that if the pseudogap arises from pairing then SC fluctuations and associated Nernst and diamagnetic responses should persist all the way up to pseudogap temperature~\cite{Lee2006}.  

 Several theoretical works in the past have studied the Nernst effect and diamagnetism in the models of SC fluctuations in various parameter regimes. One of the pertinent issues in this context is the relative importance of amplitude and phase fluctuations of the SC order parameter $\psi=\Delta e^{i\phi}$ and the role of vortices in the observed signal. Microscopically, the effect of Gaussian fluctuations around the BCS state has been investigated\cite{Ussishkin} near $T_c$ and, more recently, over a broad range of temperature and magnetic field \cite{Michaeli2009}. Also, fluctuations beyond that of the BCS paradigm have been studied using a more phenomenological description \cite{Levchenko2011}. The thermo-electric response has also been calculated in numerical simulations \cite{Mukerjee} capturing fluctuations beyond the Gaussian level through a Ginzburg-Landau functional, modeling overdoped cuprates.              
 
 Other complementary theoretical works, more relevant for the underdoped region, have utilized a `phase-only' description, by studying a two-dimensional (2D) XY model and its variants via various numerical \cite{Podolsky,Raghu2008} and analytical methods \cite{Oganesyan,Benfatto2007,Wachtel2014}. Such a description for underdoped cuprates is based on the phase-fluctuation scenario \cite{Emery1995}, where, unlike in BCS theory, $T_c$ is controlled by the superfluid density $\rho_s$ rather than the pairing gap scale $\Delta\gg \rho_s$. As a result, superconductivity gets destroyed at $T_c$ by strong phase fluctuations in the underdoped regime, whereas local pairing survives up to a much higher temperature scale $\propto \Delta$. The importance of phase fluctuations has also been emphasized in the analysis of diamagnetism \cite{Li_2010} and the Nernst effect \cite{Wang_2006} by Ong and co-workers, but this interpretation of the Nernst data has been challenged recently \cite{Chang_2012}. However, there is considerable independent evidence in underdoped cuprates that the SC order is destroyed by phase-disordering \cite{Uemura1989,Emery1995,Broun2007,Hetel2007} rather than a gap collapse. This is also what one expects in a doped Mott insulator \cite{Lee2006,Paramekanti2001}.

Studies of fluctuation diamagnetism \cite{Podolsky,Oganesyan,Benfatto2007,Wachtel2014,Raghu2008} based on the phase-fluctuation scenario have mostly ignored the effect of amplitude fluctuations and thus are constrained to describe only the extreme underdoped part of the cuprate phase diagram. A complete theoretical calculation of either diamagnetism or the Nernst effect based on a single model of SC fluctuations over the entire range of experimentally realized doping has so far not been performed.

 In this work, we aim to address the above issue and calculate the fluctuation diamagnetism (an equilibrium property, unlike the Nernst effect which is a consequence of nonequilibrium transport) based on a recently proposed phenomenological model of high-$T_\mr{c}$ superconductors \cite{Banerjee_1}. This model is of the Ginzburg-Landau (GL) type with doping and temperature dependent coefficients and is motivated by a large amount of spectroscopic data obtained from the cuprates  such as ARPES\cite{Damascelli2003}, STM \cite{Fischer2007} as well as thermodynamic and transport measurements\cite{Timusk1999}. The starting point of our description is the premise that to understand several aspects of their phenomenology, the free energy of cuprate superconductors can be expressed solely as a functional of the complex pair amplitude (e.g.~for investigating the role of pairing fluctuations in diamagnetic response), other degrees of freedom e.g. electrons, CDW being not explicit. 
 
\begin{figure}
\begin{center}
\includegraphics[height=6cm]{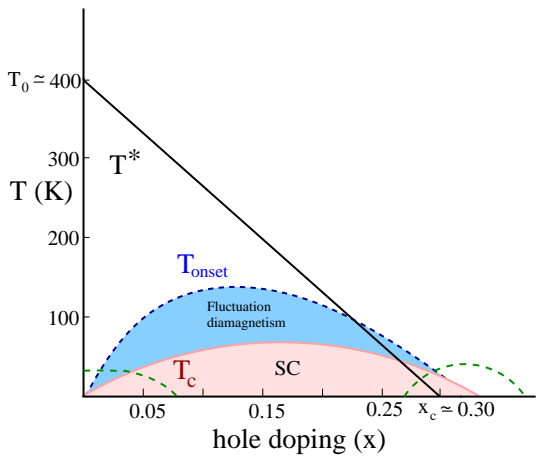}
\end{center}
\caption{ A schematic phase diagram for the model [Eq.~\eqref{Eq.functional}] in the hole doping $x$ and temperature $T$ plane. The local pairing temperature scale $T^*(x)$,  (solid black line) is an input to our phenomenological model [Eq.~\eqref{Eq.functional}] and it mimics the experimental pseudogap temperature scale. The model reproduces a dome-shaped superconducting region (shaded in pink). The region of enhanced fluctuation diamagnetism (shaded in blue) in the pseudogap phase and corresponding onset temperature $T_\mr{onset}$ are shown. The two arcs shown by dotted lines denote regions where quantum fluctuation effects, as well as other low-energy degrees of 
freedom, such as electronic and spin plus their coupling with pair degrees of freedom, need to be explicitly included in 
the free energy functional.}
\label{fig.PhaseDiagram}
\end{figure}

 There are two main inputs to our phenomenological theory\cite{Banerjee_1} -- (i) a pairing temperature scale $T^*(x)$ below which local pairing amplitude becomes substantial and (ii) the superfluid density $\rho_s(x)\propto x$, that linearly increases with doping for small x, as implied by well-known Uemura correlation \cite{Uemura1989}. As shown in the schematic phase diagram for our model in Fig.~\ref{fig.PhaseDiagram}, the temperature scale $T^*(x)$ mimics the doping dependence of the pseudogap temperature scale \cite{Timusk1999} and is much larger than $\rho_s$ in the underdoped regime. The model effectively interpolates between a phase-only description in the extreme underdoped side to a conventional Ginzburg-Landau model in the overdoped region and also accesses the intermediate regime around optimal doping.  The model is also able to produce several other experimentally observed properties of the cuprates, such as the doping dependence of the phase stiffness or $\mr{T_c}$ and the fluctuation specific heat \cite{Banerjee_1} and when coupled to nodal quasiparticles can produce Fermi arcs with details that agree well with experiments \cite{Banerjee_2}. A summary of the model and the parameters occurring in it is provided in Appendix A. 
 
 We note that there is considerable evidence for other ordering tendencies in the cuprates, e.g., nematic\cite{Kivelson1998}, stripes\cite{Kivelson2003}, checkerboard\cite{Hoffman1995}, circulating current \cite{Fauque2006} and charge density wave (CDW) \cite{Fradkin2014}. The strength and significance of each varies with the material, doping and temperature. Recently detection of short-range CDW order by X-ray \cite{Achkar2012,Chang2012,Ghiringhelli2012} and NMR measurements \cite{Wu2011} in the pseudogap and superconducting states of underdoped cuprates has attracted a lot of attention. In zero magnetic field, short-range CDW order competes with superconductivity and seems to become long-ranged only at high fields, presumably giving rise to Fermi surface reconstruction, as suggested by quantum oscillation experiments \cite{Leyraud2007}. Motivated by these findings a phenomenological nonlinear sigma model (NLSM), in terms of a coupled SC and CDW order parameter has been proposed recently~\cite{Hayward_Science,Hayward_PRB}. Subsequent work ~\cite{Wachtel2015} has pointed out that the observed magnetization and Nernst effect could arise primarily from the vortex physics that results after the charge ordering degrees of freedom have been integrated out. Further, it also attributes the rise of the X-ray structure factor, seen in experiments, to a proliferation of vortices. Our theory is similar in spirit to these ideas and the parameters in our GL-like free energy functional can be thought to be renormalized values of those that occur in a theory with a larger order parameter space after the other orders have been integrated out.
 
In this work we obtain the magnetization by performing classical Monte-Carlo (MC) simulations on the GL-like model described above. Calculations of the Nernst effect will be reported elsewhere \cite{Sarkar}.  As our model effectively reduces to a phase-only description on the underdoped side for the temperature range $T_c<T\aplt T^*$, vortices play a major role in determining the Nernst and diamagnetic responses in this regime. The vortex core energy is one of the important quantities in this regard ; we calculate its doping dependence from our model and show that it is consistent with the available estimates~\cite{Wachtel2014} for underdoped cuprates.

Our main results could be summarized as follows --\\
1. We calculate the diamagnetic response over the entire doping-temperature phase diagram for a range of magnetic fields and show that our results match well with the available experiments.\\
2. We obtain a region of enhanced fluctuation diamagnetism in the pseudogap phase extending far above the transition temperature $T_c$ and show that the boundary of the region, namely the onset temperature $T_\mr{onset}(x)$, follows a dome shaped curve tracking $T_c$ as a function of doping (Fig.~\ref{fig.PhaseDiagram}). 

 The paper is organized as follows. In Sec.~\ref{sec.GLfunctional}, we provide a brief introduction to the model employed and describe the details of the calculation of the magnetization in Sec.~\ref{sec.mag} with Sec.~\ref{sec.method} outlining the methodology of the numerical simulation. We then present our results for magnetization and vortex core-energy in Sec.~\ref{sec.Results} with a discussion in Sec.~\ref{sec.disc} and a brief conclusion in Sec.~\ref{sec.conc}. In Appendix \ref{app.Parameters}, we also describe in detail the various features of the model and parameters in it along with a summary of how the cuprate dome can be obtained from it. A discussion of the effects of quantum phase fluctuations and other competing orders on the superconducting dome is given in appendix \ref{app.TcQuantum}.

\section{Model}\label{sec.GLfunctional}
In our model, the highly anisotropic cuprate superconductor is modeled as a weakly coupled stack of $\mathrm{CuO_2}$ planes. As a first approximation, we ignore the interplane coupling. The free energy functional~\cite{Banerjee_1} $\mathcal{F}$ is defined on the $\mr{CuO}_2$ planes of the superconductor. The model describes the free energy as a functional in which the effects of the fluctuations of the order parameter phase and magnitude are coupled, and the relative importance of the latter increases with hole doping $x$. The pairing field $\psi_m=\Delta_m \exp(i\phi_m)$, with amplitude $\Delta_m$ and phase $\phi_m$, is defined on the sites $m$ of a square lattice. Microscopically, the field $\psi_m$ is expected to be related to the complex spin-singlet pairing amplitude defined on the Cu-O-Cu bonds \cite{Banerjee_1,Baskaran1988}. The functional $\mc{F}=\mc{F}_0+\mc{F}_1$ is defined as
\begin{subequations}\label{Eq.functional}
\begin{eqnarray}
&&\mathcal{F}_0(\{\Delta_m\})=\sum_m \left(A\Delta_m^2 + \frac{B}{2}\Delta_m^4\right),\\
&&\mathcal{F}_1(\{\Delta_m,\phi_m\})=-C \sum_{\langle mn\rangle}  \Delta_m \Delta_n \cos(\phi_m-\phi_n),~~~~~~
\end{eqnarray}
\end{subequations}
where $\langle mn\rangle$ represents pairs of nearest neighbour sites. The details of the choice of parameters for a specific cuprate material, e.g.~Bi2212, that we study here, are discussed in Appendix \ref{app.Parameters}. We note that the two main inputs from cuprate phenomenology are the doping and temperature dependence of the parameters $A$ and $C$. $B$ is assumed a doping independent
positive number. 
 
\section{ Magnetization}\label{sec.mag} 
The effect of a magnetic field is incorporated in our model through a bond flux $A_{mn}$, which modifies $(\phi_m-\phi_n)$ to $(\phi_m-\phi_n-A_{mn})$ in $\mathcal{F}_1$ [Eq.\eqref{Eq.functional}]. Assuming extreme type-II (infinite penetration depth) limit as appropriate for the cuprate superconductors, the magnetic field $H$ is given by the condition $\sum_{\Box} A_{mn} = \Phi$, where $\sum_{\Box}$ is a sum over a plaquette of the lattice and $\Phi=Ha^2/\Phi_0$ is the magnetic flux per plaquette in units of the universal flux quantum $\Phi_0$. As mentioned earlier the lattice constant $a$ introduces a field scale $H_0=\Phi_0/(2\pi a^2)$, which, in principle, can be deduced by comparing our results with experimental data. Here we study the phase diagram as a function of $H/H_0$. 

Introducing a magnetic field  gives rise to a diamagnetic moment ${\bf M}$ in the system. The principal goal of our calculation is to find ${\bf M}$ as a function of $T$, $H$ and $x$. To achieve this, we first define the diamagnetic current along a bond between nearest neighbour sites $m$ and $n$ as 
\begin{eqnarray}
j_{mn} = -\frac{\partial \mathcal {F}}{\partial A_{mn}}=C\Delta_m\Delta_n\sin(\phi_m-\phi_n-A_{mn}) \label{Eq.Current}
\end{eqnarray}
 ${\bf M}$ is then obtained from ${\bf j} = \boldnabla {\bf \times} {\bf M}$ by an appropriate integration. To this end we perform our calculations in a cylindrical geometry with periodic boundary conditions along the $x$ and zero current conditions along the $y$ direction. The magnetic field is radially outwards which yields a magnetization ${\bf M}$ along the same direction. We work in the Landau gauge where the $A_{mn}$ are non-zero only for $\langle mn\rangle$ along the $y$ direction and ${\bf j}$ is only along the $x$ direction. We integrate the current along the $y$ direction from the edge of the sample to the middle to obtain $M$. We use Metropolis sampling for the MC simulations. We have verified that our results are independent of gauge choice and also boundary conditions.

We would like to emphasize that there is no Meissner effect in our system since it is 
strictly two dimensional. Thus, the expression for the current we use is different from the standard London expression $\mathbf{J} \propto \mathbf{A}$, which holds
only for the so-called Coulomb gauge ($\boldnabla{\bf .}{\bf A} = 0$) and where the complex
superconducting order parameter is spatially uniform. Here, we are concerned with diamagnetism at field scales for real materials much higher than $H_{c1}$, where the Meissner effect sets in. In our numerical grid, the direction of the current is always in the periodic direction for all choice of gauges as can be seen 
from the gauge invariance of the expression for the current Eq.\eqref{Eq.Current}. Gauge invariance here is invariance under the simulatenous transformations ${\bf A} \rightarrow {\bf A} +
\nabla f_i$ and $\theta_i \rightarrow \theta_i + f_i$ for an arbitrary function $f_i$ of $i$.  Note that such a transformation is
not allowed if the $\Delta$ is spatially uniform (a necessary condition for the London expression for the current). As mentioned earlier, we have verified the gauge invariance of our results by choosing different gauges for ${\bf A}$ in our simulation.

\section{Simulation Methodology}\label{sec.method}
We perform our simulations using the standard Metropolis Monte-Carlo scheme. For obtaining the superfluid density $\rho_s$, we perform the simulations on a square lattice of size 100$\times$100 with periodic boundary conditions along both directions. At each value of doping $x$ we let the system equilibrate for about $10^5$ MC steps per site and then average over $4\times 10^5$ MC steps. As mentioned in section \ref{sec.superconductingTc} we determine the actual transition temperature  $T_c$ accurately by employing the standard finite-size scaling analysis~\cite{Weber_1987} of KT transition for small system sizes.

To calculate the magnetization we perform our simulations on a cylindrical grid with periodic boundary conditions in one direction and zero current conditions along the other. 
The magnetic flux is in the the radial direction of the cylinder with uniform flux per plaquette. The allowed values of flux are in the radial direction and  determined by the condition of zero flux in the axial direction. The current is in the azimuthal direction (in the direction in which we have periodic boundary conditions), is maximum at the edges and falls to zero and changes direction at the centre. As a consistency check, we calculate the magnetization at zero, $\pi$ and $2\pi$ flux/plaquette and find it to be zero as it should be. For each $(x,T,H)$, we perform $10^6$ MC steps per site for equilibration and a further $4 \times 10^6$ steps for thermal averaging for our largest system size. Even though we perform the simulation for a single 2D layer we assume that the actual 3D system is a collection of 2D layers with a negligible Josephson coupling between them. Thus, the only possible interaction between the pancake vortices is electromagnetic in nature. This type of interaction has been shown to not change the BKT universality class of the superfluid transition and gives a very small non-universal correction to the superfluid density jump~\cite{Raman_2009}. We thus also ignore the electromagnetic interaction among the vortices in different layers. The conversion from 2D magnetization to 3D magnetization involves division by an appropriate length along the $c$ axis. For Bi2212, the lattice spacing along the $c$ axis is 3.07 nm and the appropriate length is half this value $\sim 1.5$ nm~\cite{Kogan,Oganesyan}. The dimensionless temperature is converted into Kelvin by multiplying with $T_{0}$ which is suitably chosen for Bi2212 (see Appendix \ref{app.Parameters}). 

\section{Results}\label{sec.Results}
In this section we report our results for fluctuation diamagnetism. Based on these, we analyse the superconducting fluctuation regime in the pseudogap phase. We also discuss the role of vortex core-energy in our model.  
\subsection{Fluctuation diamagnetism}

\begin{figure}[htps]
\includegraphics[height=4.5cm,clip=]{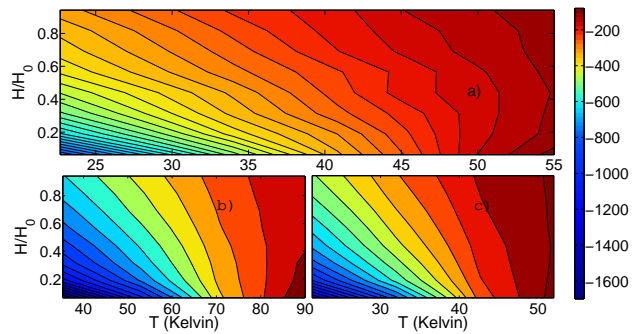}
\caption{Color map plots  of a) UD($x=0.05$), b) OPT($x=0.15$), c) OD($x=0.25$). The color bar shows the values of calculated magnetization in A/m in the field temperature ($H-T$) plane. The numerically obtained $\mr{T_c{'s}}$ are 45 K, 71 K and 42 K for UD, OPT and OD cases respectively.}
\label{Fig:magcolor}
\end{figure}

 We have obtained the values of magnetization as a function of temperature, doping and field in our model. The overall features of diamagnetism over the phase diagram are summarized in Fig.\ref{Fig:magcolor} through color map plots of the strength of diamagnetic signal as a function of $T$ and $H$ for three different values of doping from underdoped to overdoped. Having already fixed the parameters of our model, we convert the magnetization to physical units as mentioned above. The magnetic field also could be converted to Tesla by a suitable choice of the field scale $H_0$ or equivalently the coarse-graining length $a$ in our model. We can achieve good qualitative and reasonable quantitative agreement with the experimental data of Li {\em et al.}~\cite{Li_2010} with magnetization values at worst being within a factor of 2 with the measured values by choosing $H_0\approx 30-50$ T, i.e. $a\approx 25 -30~\AA$.  
 
\begin{figure}[htps]
\begin{center}
\includegraphics[height=6cm,clip=,width=9.5 cm,clip=]{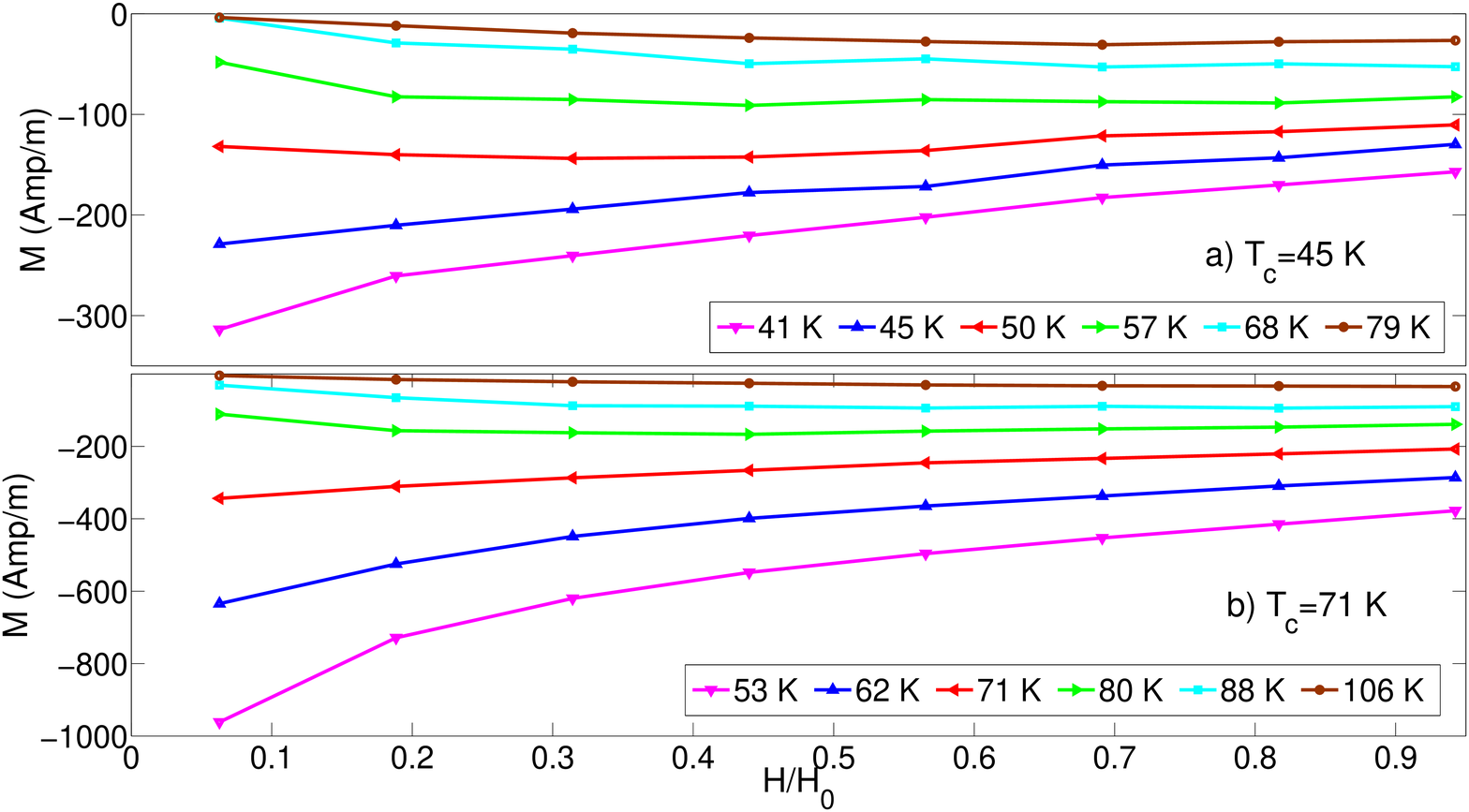}
\end{center}
\caption{ The diamagnetic response with $\mr{H/H_0}$ for a) UD ($x=0.05$) b) OPT ($x=0.15$). Our plots are in good qualitative agreement with the experimental data of Li {\em et al.}~\cite{Li_2010} for Bi2212 and are at worst within a factor of 2 of the measured values.}
\label{Fig:magvsf}
\end{figure}

 Fig.~\ref{Fig:magvsf} shows the magnetization as a function of magnetic field for $x=0.05$ and $x=0.15$ at different temperatures. The main qualitative observation that one can immediately make is that as the field decreases the magnetization appears to go to zero for temperatures $T>T_\mr{c}$ and to diverge for $T<T_\mr{c}$. This is consistent with the predictions of a renormalization group calculation in the vicinity of the BKT transition~\cite{Oganesyan}. As mentioned earlier, we do not have a Meissner effect in our calculations. The full Meissner effect diamagnetic signal for real materials is much larger than the values of magnetization that we obtain here. For the range of magnetic fields we apply, it is of the order of $10^6$ A/m per Tesla. Since this is several orders of magnitude larger than the fluctuation diamagnetic response we calculate (about 1500 A/m at its largest), we do not show it in the plots with the experimental data. Our assumption of working with a strictly two dimensional system is valid only for fields substantially larger than $H_{c1}$, which holds for the specific range of fields for which we perform our calculations. 
 
 In the following we analyse the temperature dependence of the magnetization above $T_c$ in more detail and discuss the onset temperature $T_\mathrm{onset}$ for fluctuation diamagnetism in the pseudogap state.

\subsection{The `onset' temperature for fluctuation diamagnetism}

\begin{figure}[htps]
\centering
\includegraphics[height=4.5 cm,clip=]{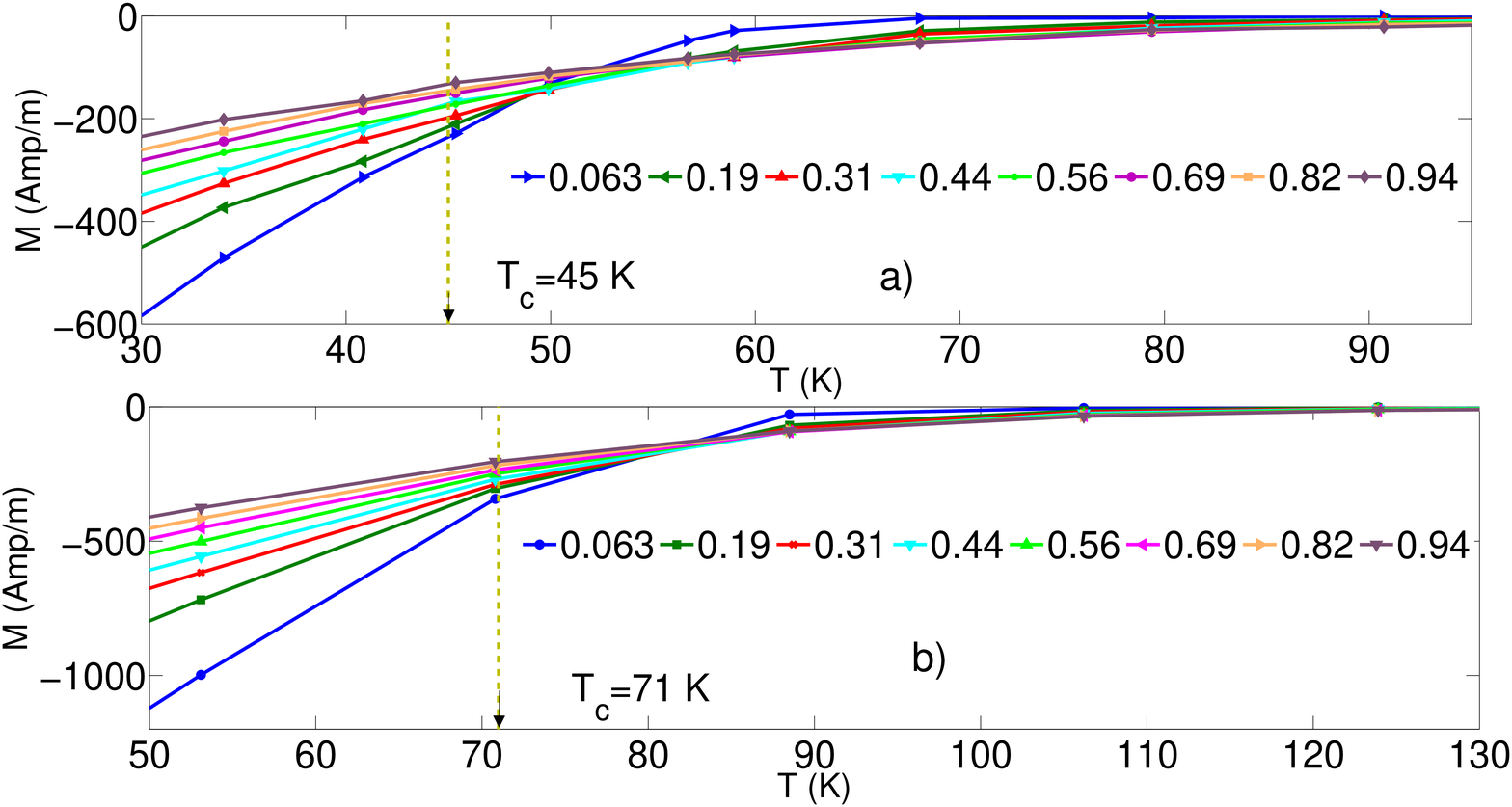}
\caption{The temperature dependence of magnetization for a) UD ($x=0.05$) and b) OPT ($x=0.15$) for different $\mr{H/H_0}$ obtained from our calculations. We can see a large diamagnetic signal above $T_c$ consistent with experiments~\cite{Li_2010,Wang_2005}}.
\label{Fig:magvsT}
\end{figure}

\begin{figure}[htps]
\centering
\includegraphics[height=8cm,width=11 cm]{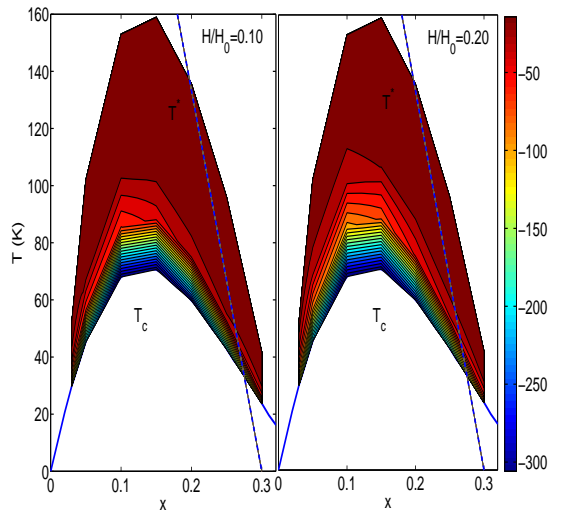}
\caption {The diamagnetic signal in the $x-T$ plane for two different values of $\mr{H/H_0}$. It can be clearly seen that diamagnetic signal follows the superconducting dome. The magnetization at these fields is obtained by interpolating between our numerical data points for fixed values of fields.}
\label{Fig:magxt}
\end{figure}

In Fig.~\ref{Fig:magvsT}, we show plots of the magnetization as a function of temperature for two different values of doping. It is evident that there is a significant diamagnetic signal persisting much above $T_c$ as seen in experiments~\cite{Li_2010}. To clearly demonstrate this point, we have plotted the magnetization above in the $x$-$T$ plane for $T>T_c$ for two values of magnetic field as in Fig.~\ref{Fig:magxt}. The diamagnetic signal is found to extend till a temperature  which is weakly dependent on the field and approximately scales as $\sim 1.5 T_c$ for the particular choice of parameters here.

Experimentally, both the Nernst effect~\cite{Wang_2006} and diamagnetism~\cite{Li_2010} have been seen to track the superconducting dome. The persistence of the Nernst and diamagnetism signal over a dome-shaped region above $T_c$, instead of the entire pseudogap state till $T^*(x)$, has been argued as evidence against the pairing origin of the pseudogap line. This is due to the expectation that if the pseudogap line is related to pairing then superconducting fluctuations should continue till $T^*$. However, on the basis of our results, we can argue that this expectation is not justified since pairing fluctuations identifiable as superconducting fluctuations, e.g.~those detected through Nernst effect or diamagnetism, are mainly controlled by $\rho_s$ or the $T_c$ scale. The GL-like model we study has the pseudogap temperature $T^*(x)$ explicitly set as the local pairing scale by construction, but, even then, the diamagnetic signal tracks the superconducting $T_c$ that is governed by the superfluid density rather than the pairing scale $T^*$ on the underdoped side.

 The putative boundary of the region of substantial fluctuation diamagnetism can be defined as the onset temperature $T_\mathrm{onset}(x)$. Experimentally the onset temperature is inferred from both Nernst~\cite{Wang_2006} effect and diamagnetism~\cite{Li_2010} measurements. In the former case $T_\mathrm{onset}$ is defined as the temperature at which the measured Nernst signal starts deviating from the high-temperature background quasiparticle contribution and, in the latter case, as the temperature where the magnetization starts decreasing rapidly away from a weakly $T$-dependent paramagnetic Van Vleck signal. In our model $T_\mathrm{onset}$ can be deduced by defining, albeit in an ad hoc manner, a threshold value of the magnetization. However, more concretely, a good qualitative measure of the onset temperature can be obtained from our model by estimating the transition temperature $T_c^\mathrm{mf}$ via a single-site mean-field approximation~\cite{Banerjee_1}. $T_c^\mathrm{mf}(x)$ gives a measure of the temperature scale corresponding to the local superfluid density and is found to be approximately $1.5T_c$. Fluctuation effects destroy the global phase coherence and reduce the mean-field transition temperature from $T_c^\mathrm{mf}$ to the actual transition temperature $T_c$. However, one would expect a manifestation of substantial SC fluctuations, such as fluctuation diamagnetism, to persist over a temperature range $T_c<T<T_c^\mathrm{mf}\approx T_\mathrm{onset}$. These considerations lead to the schematic phase diagram of Fig.~\ref{fig.PhaseDiagram}.
  
\begin{figure}[htps]
\centering
\includegraphics[height=6cm,clip=]{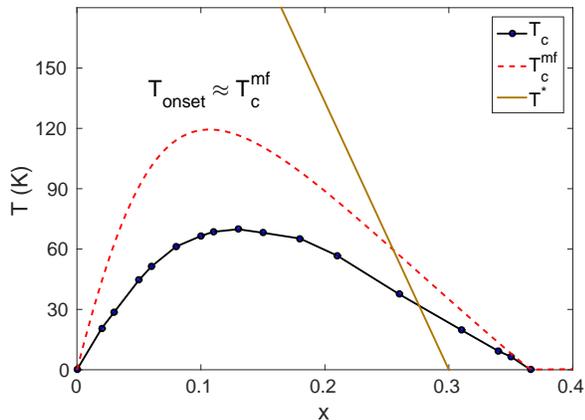}
\caption{The estimate of transition temperature $T_c^\mathrm{mf}(x)$ in single-site mean field approximation~\cite{Banerjee_1}. $T_c^\mathrm{mf}$ can be defined as a measure of the onset temperature $T_\mathrm{onset}$.}
\label{Fig:Tc_MFT}
\end{figure}

\subsection{Phase and amplitude fluctuations}
\begin{figure}[htps]
\begin{center}
\begin{tabular}{c} 
\includegraphics[height=6cm,width=9.5 cm]{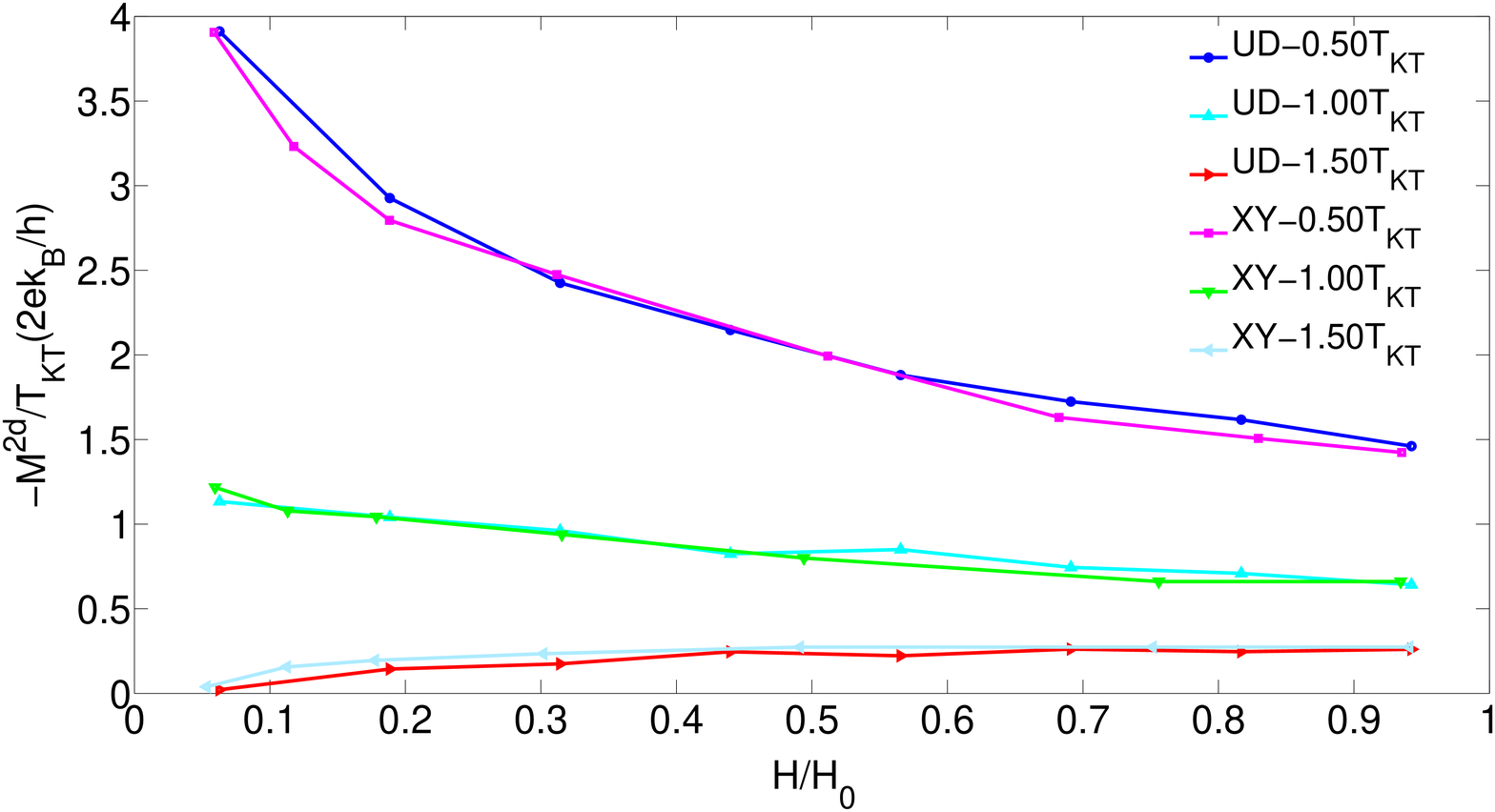}
\end{tabular}
\end{center}
\caption{Our numerically calculated 2D magnetization $\mr{-M^{2d}}$ scaled by our obtained $\mr{T_{KT}}$ for UD (x=0.05) overlaps with the $\mr{-M^{2d}/T_{KT}}$ results obtained by Podolsky et al.~\cite{Podolsky} in XY model at three different temperatures $\mr{T<T_{KT}}$, $\mr{T=T_{KT}}$ and $\mr{T>T_{KT}}$. It essentially reveals  that in our extreme UD region amplitude $\Delta$ gets frozen and the region is effectively described by `phase only' model.}
\label{Fig.Mag_2dUD}
\end{figure}

The suppression of $T_\mr{c}(x)$ relative to $T^*(x)$ is an indicator of the strength of superconducting fluctuations. The fact that $T_\mr{c}(x)$ is dome shaped while $T^*(x)$ decreases monotonically with increasing $x$ shows that fluctuations get weaker with increasing doping. The two extremes of the strength of fluctuations are represented by the phase-only $XY$ model (`strong fluctuations') and the Gaussian model (`weak fluctuations'). The magnetization in the former model has been calculated and found to be in reasonable agreement with experimental data~\cite{Podolsky}. Our data for extreme underdoped samples are in agreement with the results of these calculations as shown in  Fig~\ref{Fig.Mag_2dUD}.

The strong fluctuation limit corresponds to the fluctuations essentially in the phase $\phi$ of the superconducting order parameter with the amplitude $\Delta$ being frozen. As $x$ increases, amplitude fluctuations start becoming more significant even as the overall strength of fluctuations decreases till one arrives at regime where the fluctuations are Gaussian and cannot be divided into contributions from amplitude and phase in any meaningful sense. As can be seen from Fig.~\ref{Fig:magvsf}, our calculations show that the qualitative behavior obtained from the $XY$ model persists even when amplitude fluctuations develop changing only the overall magnitude of the magnetization.

\subsection{Vortex core-energy and superconducting fluctuations}
 
In this section we analyse the doping dependence of vortex core-energy in our model~\cite{Banerjee_1}. As we discuss here, vortex core-energy has important consequences for the fluctuation regime above $T_c$, especially in the underdoped side.

We use the free-energy functional of Eq.\eqref{Eq.functional} to find the core energy  of vortices at $T=0$. We expect the core-energy to be weakly temperature dependent for the underdoped side in the temperature range of interest here. To generate a single vortex configuration we minimize $\mathcal{F}$ with respect to $\Delta_m$ and $\phi_m$ at each site while keeping the topological constraint of total $2\pi$ winding of the phase variables at the boundary of a $N\times N$ lattice. This is a standard way of obtaining a vortex configuration of vorticity $k=1$ with the vortex core at the middle of the central square plaquette in the computational lattice. In this manner we obtain the optimal energy $E_v$ of a vortex for system of size $N\times N$.

\begin{figure}
\begin{center}
\begin{tabular}{c} 
\includegraphics[height=6cm]{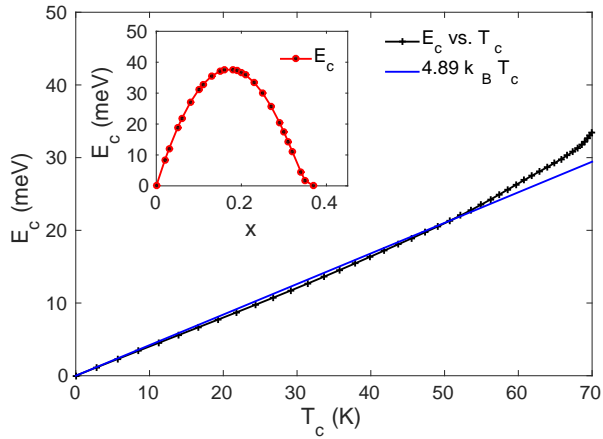}
\end{tabular}
\end{center}
\caption{Vortex core energy $E_c$ as a function of the transition temperature $T_c$ in the underdoped side. The core-energy is found to be $\approx 5T_c$ for small $x$. The inset shows $E_c(x)$ over the entire doping range.}
\label{fig.VortexCoreEnergy}
\end{figure}

The core energy $E_c$ of a single vortex is naturally described as the extra energy $\Delta E_v=E_v-E_0$ where $E_0$ is the energy of the ground state configuration and $E_v$ is the total energy of a single vortex configuration, from which the elastic energy due to phase deformation is subtracted, i.e.
 \begin{eqnarray}
\Delta E_v&=&E_c+\pi \rho_s(0) \ln(R/l)
\end{eqnarray}
The quantity $R$ is defined as $R=(N-1)a/\sqrt{\pi}$ so that $\pi R^2$ is the area of the computational lattice. 
We have estimated $E_c$ from the intercept of the $\Delta E_v$ vs. $\ln(R/l)$
(different system sizes) straight line~\cite{Banerjee_1}. As shown in Fig.~\ref{fig.VortexCoreEnergy} (inset) the core energy $E_c$ follows a dome shaped curve as a function of $x$. In the underdoped side, $E_c$ is found to scale linearly with $T_c$. Such a scaling in our phenomenological description is in conformity with the phase-fluctuation scenario \cite{Emery1995} and the idea of cheap vortices \cite{Lee2006}.
 
 In a recent work Wachtel et al. have developed~\cite{Wachtel2014} a vortex-only description, presumably applicable to underdoped cuprates, to calculate the magnetization and Nernst coefficient above $T_c$. For low fields, they obtain the magnetization $M\simeq -T H/\Phi_0^2n_f$ and the Nernst coefficient $\alpha_{xy}\simeq (E_c/T)(cM/T)$, where $n_f$, the density of free vortices, is controlled by the vortex core-energy, i.e.~$n_f\propto 2 e^{-E_c/T}$ as $H\rightarrow 0$. Based on our model and the calculated vortex core-energy a similar vortex-only description could also be obtained for the underdoped side where our model effectively reduces to a XY model. By fitting temperature dependence of $\lim_{H\to 0} \alpha_{xy}/H$ with their model Wachtel et al. found the core-energy $E_c\approx 4-5 T_c$. As shown in Fig.~\ref{fig.VortexCoreEnergy}, we also obtain similar ratio for $E_c/T_c$ for small $x$.

\section{Discussions} \label{sec.disc}

\subsection{GL-like functional and the upper critical field}

An important and rather controversial issue of recent interest is the value and the doping dependence of the upper critical field $H_{c2}$ in the cuprates. One of the experimental methods to determine $H_{c2}$ involves extrapolation of the measured Nernst signal of Bi2212 at $T_c$ and obtaining the $H_{c2}$ from a scaled plot of Bi2201, for which the putative $H_{c2}$ can be directly accessed. This approach has been advocated by Ong and coworkers~\cite{Ong_2003, Wang_2006} who reported the increase of $H_{c2}$ with underdoping. Other approaches utilizing either analysis of the magnetoconductivity~\cite{Ando_2002,Albenque_2011} of YBCO above $T_c$ or obtaining a characteristic field $H^*$ from the peaks of the Nernst signal versus magnetic field isotherms of Eu-LSCO cuprate and fitting $H^*$ to a Gaussian fluctuation form~\cite{Chang_2012} give a completely different dependence where $H_{c2}$ decreases with underdoping. The values of $H_{c2}$ obtained from the two methods can differ by as much as a factor of 2 in the UD region. 

In our coarse-grained model, we assume the lattice spacing $a\apgt \xi_0$, the zero-temperature coherence length and hence focus on low fields $H<H_0\aplt H_{c2}$. Nevertheless, it is worthwhile to mention that one can take a continuum limit of the model [Eq.\eqref{Eq.functional}] and deduce that the coherence length $\xi_0$ goes as $\sqrt{x}$ for small $x$ due to the fact that the coefficient $C\sim x$. This would suggest $H_{c2}(T=0)\sim 1/x$. We note that this is indeed the rough $x$ dependence of this quantity obtained by Ong et al.~\cite{Ong_2003}. However, the validity of the above mentioned continuum limit of Eq.~\eqref{Eq.functional} to the scale of $\xi_0$, which could be order of a few Cu-Cu lattice spacing in the underdoped side, is not entirely clear and one needs more microscopic considerations to settle this issue.    

\subsection{Pseudogap and Competing orders}\label{subsec.Competing}

Our free energy functional only contains the pairing order parameter. Since there are experimental evidences of the presence of other kinds of order in the 
cuprates, it is natural to ask how reliable our model is. The point we would like to make is that we seek to elucidate the role of only the superconducting fluctuations in the phenomenology of the cuprates. To that extent, we work with a model which only contains superconductivity and no other types of order. As mentioned in the introduction, the model can be thought of as arising from one with other types of order integrated out. 

Nevertheless, an important question is how much of the phenomenology of the pseudo gap can be attributed to the presence of orders other than superconductivity. It appears that the evidence is not sufficiently compelling yet as to ascribe the pseudgap entirely to some order competing with superconductivity. While a large number of ordering
tendencies, e.g., orbital current~\cite{Fauque2006}, spin density wave (SDW) \cite{Fradkin2014}, charge density wave (CDW) \cite{Achkar2012,Chang2012,Ghiringhelli2012} etc., have been detected experimentally, their explicit role in the origin of the pseudogap is still not very well understood. For example, it is not clear how the orbital current order, detected by polarized neutron scattering~\cite{Fauque2006}, can lead to a large ($\sim 50$ meV) pseudogap and the phenomena of SDW ordering does not seem to be ubiquitous in all cuprates~\cite{Kivelson2003,Fradkin2014}. 

 Recent experiments~\cite{Achkar2012,Chang2012,Ghiringhelli2012} have detected strong CDW correlations in several cuprates. But the CDW is at best a short-rang order with a small correlation length in zero field and only becomes long ranged at high fields and
low-temperature \cite{Chang2012,Wu2011}. The interplay between superconductivity and 
short-range CDW order can be studied within our framework by incorporating additional terms for the CDW order parameter. This would be similar in 
spirit to a recent work~\cite{Hayward_Science,Hayward_PRB,Allais}, where the 
effect of fluctuating CDW order in an addition to superconductivity has been taken into account in terms of an expanded $O(6)$ order parameter. To our understanding, one of the conclusions of the above mentioned study is that
the fluctuating CDW, due to its short correlation length, does not
significantly influence the pseudogap and other related features seen in
ARPES \cite{Allais} or low-field diamagnetism\cite{Hayward_PRB}, as obtained solely from pairing fluctuations. This validates our approach of retaining only superconducting fluctuations
to study diamagnetism.

\section{Conclusion}\label{sec.conc}

 We have used a phenomenological Ginzburg-Landau-like energy functional for the superconducting order parameter which allows us to determine the doping dependence of diamagnetism in the cuprates in addition to its dependence on temperature and magnetic field. We find that our results are in good qualitative  and quantitative agreement within a factor of 2 of experimental data obtained on Bi2212~\cite{Li_2010,Li_2013}. We show that the diamagnetic response as a function of doping tracks the superconducting dome whose scale is set by $T_c$ and not the pairing scale, which is the pseudogap temperature in our model. This leads to a scenario where substantial local pairing can survive till the pseudogap temperature in the underdoped cuprates, even though superconducting fluctuations as manifested in diamagnetic response only exist up to a much lower temperature. 

\section{Acknowledgements} 
K.S. would like to thank CSIR (Govt. of India) for support. S.B. acknowledges the support of DOE-BES DE-SC0005035 grant. S.M. thanks the DST (Govt. of India) for support. T.V.R. acknowledges the support of the DST Year of Science Professorship, and the hospitality of the NCBS, Bangalore. The authors would like to thank Nabyendu Das, Chandan Dasgupta, Vadim Oganesyan, Daniel Podolsky and Srinivas Raghu for stimulating discussions.

\appendix

\section{The free energy functional} \label{app.Parameters}

The functional $\mc{F}=\mc{F}_0+\mc{F}_1$ is defined as
\begin{subequations}\label{Eq.functional2}
\begin{eqnarray}
&&\mathcal{F}_0(\{\Delta_m\})=\sum_m \left(A\Delta_m^2 + \frac{B}{2}\Delta_m^4\right),\\
&&\mathcal{F}_1(\{\Delta_m,\phi_m\})=-C \sum_{\langle mn\rangle}  \Delta_m \Delta_n \cos(\phi_m-\phi_n),~~~~~~
\end{eqnarray}
\end{subequations}
where the pairing field $\psi_m=\Delta_m \exp(i\phi_m)$, with amplitude $\Delta_m$ and phase $\phi_m$, is defined on the sites $m$ of a square lattice.
$\langle mn\rangle$ represents pairs of nearest neighbour sites. The coefficient of the quadratic term, $A$ is chosen to be proportional to $(T-T^*(x))$, where $T^*(x)$ is a local pairing scale which we identify with the pseudogap temperature scale \cite{Timusk1999}. The magnitude of local pair amplitude $\langle \Delta_m\rangle$ increases substantially \cite{Banerjee_1} as $T$ goes below $T^*(x)$ and $A$ changes sign. We take $T^*(x)$ to follow a simplified linear $x$ dependence, i.e.~$T^*(x)=T_0(1-x/x_c)$, mimicking the doping dependence of experimentally measured pseudogap line \cite{Timusk1999}. As shown in Fig.\ref{fig.PhaseDiagram}, $T^*(x)$ linearly decreases with $x$, going from $T=T_0$ at $x=0$ to $T=0$ at $x=x_c$. The occurrence of superconductivity, characterized by a non-zero stiffness $\rho_s$ for long-wavelength phase fluctuations, depends on the parameter $C$. We take $C\propto x$ and as a result the superconducting transition temperature calculated in our theory turns out to be proportional to $x$ for small $x$, in conformity with well-known Uemura correlation \cite{Uemura1989}. This also serves to make fluctuations at low doping easily available enabling us to produce the supercondcuting dome in the phase diagram. However, this is not the only consideration that goes into determining the form of $C$ since a similar effect can also be obtained by making $B$ doping dependent. The form of $C$ can also be motivated from microscopic considerations. In a microscopic theory, the parameter $C$ would naturally originate from the hopping amplitude of Cooper pairs between sites. If the superconducting state were to arise from doping a Mott insulator, it would be reasonable to assume that this parameter would be proportional to the doping $x$ (at least for small values), as is the case in resonating valence bond theory \cite{Baskaran1988}. 

As natural in a phenomenological theory, the parameters of the above functional are chosen to be consistent with experiment. The doping and temperature dependence of the coefficients are parametrized as $A(x,T)= (f/T_0)^2[T-T^*(x)]e^{T/T_0}$, $B=bf^4/T_0^3$ and $C(x)=xcf^2/T_0$; $f$, $b$, $c$ are dimensionless and $T^*(x)=T_0(1-x/x_c)$ with the energy scale $T_0$ and doping concentration $x_c=0.3$ controlling the pseudogap temperature scale~\cite{Banerjee_1}. The phenomenological parameters $f$, $b$, $c$ vary for different cuprates and $T_0$ is the bare pseudogap temperature extrapolated to zero doping. The exponential factor $e^{T/T_0}$ appearing in $A$ is not very crucial for the purpose of the present study in the relevant range of temperature ($\aplt T^*(x)$) of fluctuation diamagnetism. This factor suppresses average local gap magnitude $\langle \Delta_m\rangle$ at high temperatures ($T\apgt T^*(x)$) with respect to its temperature independent equipartition value $\sqrt{T/A(x,T)}$ which will result from the simplified form of the functional (Eq.\eqref{Eq.functional2}) being used over the entire range of temperature. Such a suppression is natural in a degenerate Fermi system; the relevant local electron pair susceptibility is rather small above the pair binding temperature and below the degeneracy temperature.

We show below that the forms of the parameters $A$, $B$ and $C$ specified above allow us to reproduce the superconducting dome. However, it is also important to mention that
having chosen $A$, $B$ and $C$ to reproduce the superconducting dome, the doping and temperature dependences of other physical properties like
the superfluid density, the magnitude of the local gap and the
specific heat in the presence and absence of a magnetic field, also agree
very well with experiments~\cite{Banerjee_1}. A related paper, authored by two of us,~\cite{Banerjee_2} shows that the
scattering of nodal quasiparticles by superconducting fluctuations
described by this phenomenological functional, can describe several
features of ARPES data on the cuprates including the appearance of
Fermi arcs. Thus our model is able to explain a fairly large number of observations on the cuprates based on a few phenomenological inputs.

While, underdoped phenomenology plays an important part in determining the form of our model, e.g.~to determine the doping dependence of the parameter $C$ as mentioned above, it also produces the standard GL theory for conventional superconductors on the overdoped side. The amplitude of pairing approaches zero at $T_c$, or in other words the actual or `renormalized' pairing scale $\widetilde{T}^{*}(x)\approx T_c(x)$ \cite{Banerjee_1}, on the overdoped side in our theory. This is in conformity with the common expectation that the BCS theory or mean-field GL theory is more appropriate for overdoped cuprates.

For Bi2212, which has a $T_c^\mathrm{opt}\simeq 91$ K at $x=x_\mr{opt}\simeq 0.15$, we choose $f\simeq 1.33$, $b=0.1$, $c\simeq0.3$ with $T_0\simeq400$ K. This choice of parameters leads to an optimal BKT transition temperature $T_\mathrm{KT}^{\mathrm{opt}}\approx 75 K$ for the 2D system that we study. The small but finite interlayer coupling between $\mathrm{CuO}_2$ planes is expected to lead to a somewhat higher $T_c$.

For the 2D system that we study, the superconducting transition is of the BKT type with quasi long-range order below the BKT transition temperature $T_\mr{KT}$ which we identify as the superconducting transition temperature $T_c$ in our model. In the cuprates, the small but finite inter-layer coupling between $\mr{CuO}_2$ planes is expected to lead to a slightly higher $T_c$. The interlayer coupling can be easily incorporated in our model in the manner of Lawrence and Doniach\cite{Lawrence1971}. Since this coupling is, in practice, quite small (e.g.~the measured anisotropy ratio in Bi2212 is about 100), it makes very little difference quantitatively to most of our estimates. Also, our main focus here is the region above $T_c$, and it has been shown that vortex fluctuations in 3D anisotropic XY model effectively become two-dimensional and superconducting planes to a large extent become decoupled above $T_c$ \cite{Minnhagen1991}.  
 
 Eq.\eqref{Eq.functional2} with the choice of parameters described above has been shown to lead to observed parabolic shape of $T_c(x)$ [Fig.\ref{sfldensity}(inset)] as a function of $x$ and temperature and doping dependence of various other quantities like superfluid density, the local gap magnitude $\langle\Delta_m\rangle$, the specific heat etc. in agreement with experimental results~\cite{Banerjee_1}. We discuss a few generalities of our model below, as relevant for the present context; a detailed discussion can be found in ref.~\onlinecite{Banerjee_1}.

\subsection{General aspects of the model}

As can be seen from Eqns.~\eqref{Eq.functional2}, our model has a term that is quadratic and quartic in the amplitudes and, since $\Delta_m\Delta_n\cos(\phi_m-\phi_n)=-(|\psi_m-\psi_n|^2-\Delta_m^2-\Delta_n^2)$, the term $\mathcal{F}_1$ can be readily identified with the discretized version of usual spatial derivative term $|\nabla \psi|^2$. The model is thus of the form of a Ginzburg-Landau model, albeit one that 
is defined on a lattice at the outset. The lattice here should be thought of as a phenomenological one that emerges upon coarse graining and is not the underlying 
physical lattice of the system. Nor is the lattice parameter here related to any underlying granularity of the system. Hence, our model can be thought of as the discretized version of a continuum theory with the lattice spacing $a$ as a suitable ultraviolet cutoff to describe long wavelength physics. In the presence of magnetic field $H$, the lattice constant is also equivalent to a field scale $H_0=\Phi_0/(2\pi a^2)$, defined through the flux quantum $\Phi_0=hc/2e$.  We have checked that the calculated magnetization is indeed independent of this cutoff for the relevant range of field $H < H_0$. In principle, the field scale $H_0$ can be obtained by fitting the field dependence of magnetization with that of experiment.

 The free-energy functional in Eq.~\eqref{Eq.functional2} can also be viewed as the Hamiltonian of an XY model with fluctuations in the magnitude of `planar spin' $\psi_m$, where the term $\mathcal{F}_0$ simply controls the temperature and doping dependence of the magnitude. The form of the free-energy functional might seem superficially similar to the widely used model of granular superconductors \cite{Ebner1981}. However, we would like to re-emphasize that we do not assume any underlying granularity of our system, as mentioned above.  Such phenomenological lattice models, in the extreme $XY$ limit, have been employed in the past to study superconductivity in non-granular lattice systems, especially in the context of cuprates~\cite{Kivelson_1999,Paramekanti_2000,Franz_2006,Podolsky}, as mentioned in the introduction.

Additionally, even though the form of our functional is mainly
motivated by cuprate phenomenology, it is worthwhile to mention that
a similar functional arises quite naturally in a strong correlation framework for a doped Mott insulator, see, e.g.~refs.\onlinecite{Baskaran1988,Dzrazga,Banerjee_1}. In general, in such a functional, the single-site term $\mathcal{F}_0$ will have more complicated form\cite{Dzrazga}, having many terms in a power series expansion of $\Delta_m$, in addition to the quadratic and quartic ones that our functional does. But, as we discuss below, the superconducting dome is reproduced quite reasonably by truncating the functional to quartic order. In addition, several other experimentally observed thermodynamic properties of the cuprates over the entire pseudogap regime are also reproduced by this simplified form of the functional~\cite{Banerjee_1}.

\subsection{ Superconducting Transition Temperature}\label{sec.superconductingTc}
\begin{figure}[htps]
\begin{center}
\includegraphics[height=6cm,clip=,width=9 cm,clip=]{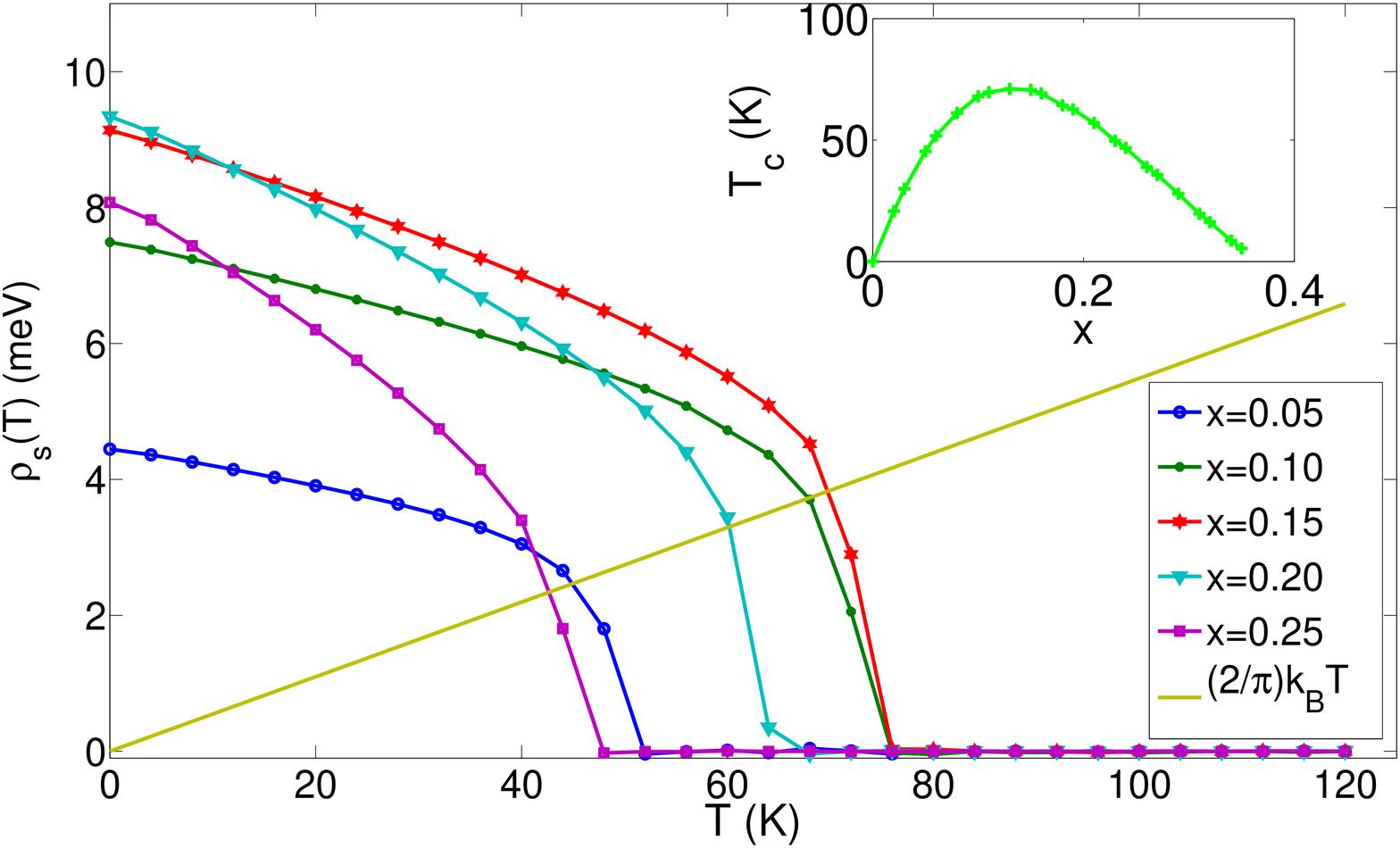}
\caption{The superfluid density for different $x$. The intersection of the straight line of slope $2/\pi$ and $\rho_s(T)$ at different $x$ gives an estimate of $T_c(x)$, the Nelson-Kosterlitz~\cite{Nelson_1977} line. $T_c$ is then determined more accurately by a finite size scaling analysis of the BKT transition~\cite{Weber_1987}. (Inset) $T_c$ as a function of $x$ from our calculations that reproduces the experimentally observed parabolic dome. (See also Appendix ).}\label{sfldensity}
\end{center}
\end{figure}

 The superconducting state is characterized by macroscopic phase coherence. For superconductivity in cuprates described by the functional [Eq.\eqref{Eq.functional}] this means a non-zero value for the superfluid stiffness or superfluid density $\rho_s(x,T)$, formally defined as $\rho_s=\frac{1}{N}(\frac{\partial^2\mc{<F>}}{\partial \theta^2})_{\theta\rightarrow 0}$, where $\theta$ is the phase twist applied along one of the two orthogonal directions and $N$ is the total number of sites. This leads to the formal expression for $\rho_s$
\begin{eqnarray}
&&\rho_s=\frac{C}{2N}\left\langle\sum_{m,\mu}\Delta_m\Delta_{m+\mu}\cos(\phi_m-\phi_{m+\mu})\right\rangle \nonumber\\
&&-\frac{C^2}{2NT}\sum_\mu\left\langle\left(\sum_{m,\mu}\Delta_m\Delta_{m+\mu}\sin(\phi_m-\phi_{m+\mu})\right)^2\right\rangle,
\end{eqnarray}
where $\mu=x,y$. We calculate the superfluid stiffness as function of doping and temperature by MC simulation of our model [Eq.\eqref{Eq.functional}] (see Section \ref{sec.method}) using the above formula and obtain the BKT transition temperature $T_\mr{KT}(x)$ accurately through the Nelson-Kosterlitz criterion~\cite{Nelson_1977} $\rho_s(T_\mr{KT})/T_\mr{KT}=2/\pi$ in conjunction with finite-size scaling analysis~\cite{Weber_1987}. The results are summarized in Fig.\ref{sfldensity}.

 As also mentioned earlier, we want to re-emphasize one point regarding our identification of $\mr{T_{KT}} \equiv \mr{T_c}$. On the basis of Kosterlitz-Thouless RG analysis Benfatto et al.~\cite{Benfatto1,Benfatto2} have demonstrated that in layered superconductors, which generally have small interlayer Josephson coupling, the Kosterlitz-Thouless {\it {behavior}}(e.g. superfluid stiffness jump at $\mr{T_{KT}}$ which evolves to a rapid turnover at $\mr{T_c \gtrsim T_{KT}}$) persists when vortex core energy is very low.

The calculated $T_c(x)$ is approximately of the same parabolic shape [see Fig.\ref{sfldensity} (inset)] as found experimentally. The reasons for the qualitative disagreement at both ends are not
difficult to understand. For very small $x$, as well as for $x$ near $x_c$ , our free-energy functional needs to be extended by including quantum phase fluctuation effects. For such values of $x$, zero-point fluctuations are important because the phase stiffness is small. The quantum fluctuations are also expected to modify the simple Uemura scaling \cite{Uemura1989} to more appropriate quantum critical scaling in the extremely underdoped cuprates \cite{Broun2007,Hetel2007}. Additionally, low-energy mobile electron degrees of freedom need to be considered explicitly for $x$ near $x_c$. We briefly discuss the role of quantum phase fluctuation effects and other possible competing orders in determining the detailed shape of $T_c(x)$ in Appendix \ref{app.TcQuantum}.

\section{Effects of quantum phase fluctuation and competing orders on $T_c(x)$}\label{app.TcQuantum}
\begin{figure}[htps]
\begin{center}
\begin{tabular}{c}
\includegraphics[height=6cm]{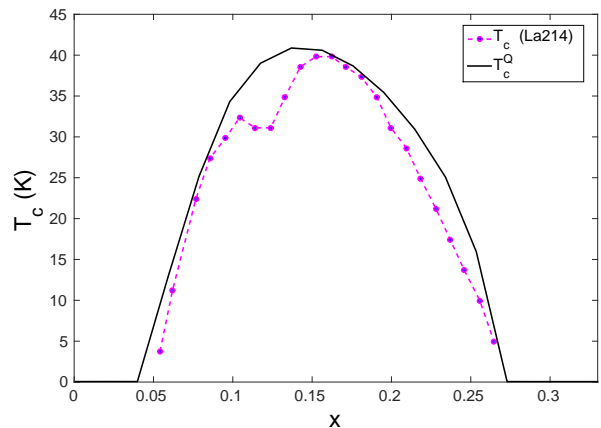}
\end{tabular} 
\end{center}
\caption{Effect of quantum phase fluctuation on $T_c(x)$ curve. A reasonably good comparison can be obtained with experimental $T_c(x)$ curve for La214 with
following choice of parameters $x_c=0.345$, $c=0.33$, $b=0.155$,
$f=1.063$ and $V_0=0.15T_0$ with $\Delta_0(x=0)=82$ meV. The dip
of the experimental $T_c$ around $x\sim0.12$ is due to the $1/8$ `stripe anomaly' \cite{Moodenbaugh}.}
\label{fig.TcQuantum}
\end{figure}

We have shown in Sec.~\ref{sec.superconductingTc} that calculated SC transition temperature $T_c$ in our model follows a dome-shaped curve as a function of $x$ (Fig.~\ref{sfldensity}). In the extreme underdoped and overdoped regimes, where the superfluid density becomes small in our model, one needs to take into account the effect of quantum phase fluctuations. These would renormalize $T_c$ to zero at finite doping in the underdoped side and their importance is well-supported by experiments~\cite{Broun2007,Hetel2007} and theoretical analysis~\cite{Franz_2006}. We can incorporate quantum phase fluctuation effects in our formalism \cite{Banerjee_1} by supplementing the free-energy functional of Eq.\eqref{Eq.functional} with the following term 
\begin{eqnarray}
\mathcal{F}_Q(\{\hat{q}_m\})&=&\frac{1}{2}\sum_{mn} \hat{q}_m V_{mn} \hat{q}_n \label{Eq.functionalQ}
\end{eqnarray}
Here $\hat{q}_m$ is the Cooper pair number operator at site $m$, and $\phi_m$ in Eq.\eqref{Eq.functional} should be treated as a quantum mechanical operator $\hat{\phi}_m$, canonically conjugate to $\hat{q}_m$ so
that $[\hat{q}_m,\hat{\phi}_n]=i\delta_{mn}$. We take the simplest possible form for $V_{mn}$ i.e.~$V_{mn}=V_0\delta_{mn}$, where $V_0$ is the strength of on-site Cooper pair interaction. We obtain \cite{Banerjee_1} a single-site 
mean field estimate of $T_c(x)$, namely $T_c^Q(x)$, including the effect of $\mathcal{F}_Q$ as shown in Fig.\ref{fig.TcQuantum}. The $T_c(x)$ dome indeed terminates at finite $x$ away from $x=0$, as seen in experiment. As an example, we show in Fig.\ref{fig.TcQuantum} that quantitative agreement for $T_c $ for a specific cuprate, $\mathrm{La_{2-x}Sr_xCuO_4}$ is possible with a particular choice of parameters. 

 We do not include these quantum fluctuations
explicitly for calculating fluctuation diamagnetism as they bring about qualitative changes only at the extreme end of the dome on the underdoped side. For other values of $x$, these fluctuations only renormalize the values
of the parameters $A$, $B$ and $C$ of our functional.
We assume that such renormalizations have already been taken into account
while choosing these parameters in tune with experiments. 

One can ask if there are other effects of quantum fluctuations that go beyond simply renormalizing parameters 
in our free energy. An example is the presence of a dip (Fig.~\ref{fig.TcQuantum}) in $T_c$ at $x=1/8$ due to concurrent stripe 
order \cite{Moodenbaugh}. The effect of such stripe order can in principle be taken into account in a multi-order-parameter functional and 
integrated out to produce our functional as has been explained earlier. The stripe order has been seen to be most dominant 
only close to $x=1/8$ and diminishing rapidly away from it~\cite{Wu2011}.  It is thus not obvious whether it would have 
any significant effects on the extreme underdoped side, such as, for instance leading to the ultimate demise of $T_c$ at $x~=0.05$, which is 
far away from x=1/8.

\end{document}